# Students' Interdisciplinary Reasoning about "High-Energy Bonds" and ATP

Benjamin W. Dreyfus, Benjamin D. Geller, Vashti Sawtelle, Julia Svoboda, Chandra Turpen, and Edward F. Redish

*Department of Physics, University of Maryland, College Park MD 20742-4111*

**Abstract.** Students' sometimes contradictory ideas about ATP (adenosine triphosphate) and the nature of chemical bonds have been studied in the biology and chemistry education literatures, but these topics are rarely part of the introductory physics curriculum. We present qualitative data from an introductory physics course for undergraduate biology majors that seeks to build greater interdisciplinary coherence and therefore includes these topics. In these data, students grapple with the apparent contradiction between the energy released when the phosphate bond in ATP is broken and the idea that an energy input is required to break a bond. We see that students' perceptions of how each scientific discipline bounds the system of interest can influence how they justify their reasoning about a topic that crosses disciplines. This has consequences for a vision of interdisciplinary education that respects disciplinary perspectives while bringing them into interaction in ways that demonstrate consistency amongst the perspectives.

## INTRODUCTION

While the treatment of energy in the traditional introductory physics curriculum focuses on mechanical energy, developing a physics curriculum for life sciences students that is intended to build cross-disciplinary coherence requires engaging with energy concepts as they are understood in biology and chemistry. A major piece of this is the energy associated with chemical bonds, which is involved in the energy transformations most relevant to biological systems. A particular molecule of interest is ATP (adenosine triphosphate), which biology students know as "the energy currency of the cell." In this study we look not only at the substantive issues involved in integrating chemical bond energy and ATP in a physics course but at the implications for interdisciplinary science education.

ATP is produced during cellular respiration and photosynthesis. In the ATP hydrolysis reaction, a bond is broken to remove the terminal phosphate group from the ATP molecule, and other bonds are formed. These new bonds have a greater bond energy (or, equivalently, they are represented by a deeper potential well), so the net result is a release of energy, which is used to power various cellular processes. As a shorthand, many biology texts and instructors refer to the phosphate bond in ATP as a "high-energy bond."

Students' conceptual difficulties with ATP and bond energy are well documented in the biology and chemistry education literature. In biology, Novick [1] and Gayford [2] both write about student confusion about "energy stored in bonds" and the misleading terminology of "high-energy bonds," particularly in regard to ATP. In chemistry, Boo [3] documents students' "alternative conception" that bond making requires energy input (even in non-biological reactions). Galley [4] also documents "exothermic bond breaking," as we will discuss at greater length. Teichert & Stacy [5] show that students (when discussing ATP) can simultaneously express the idea that energy is released when a bond is formed, and that energy is released when a bond is broken.

## METHODOLOGY

The context of this study was the pilot year of a new introductory physics course [6] for undergraduate biology students. This course is unusual in that biology and chemistry are required as prerequisites, and students are therefore expected to draw on their knowledge from these other disciplines. The course placed a particular emphasis on building coherent models of energy and thermodynamics across potential disciplinary barriers.[7] This included an extensive thread on chemical bond energy, which connected chemical bonds to electric potential energy and other constructs "native" to physics courses. Students were given a series of tasks in which they were to model chemical bonds as potential wells, and use reasoning similar to conventional conservation-of-mechanical-energy problems. This paper uses two data sources to probe students' reasoning about bond energy and ATP: we look briefly at quantitative data from a multiple-choice quiz question to obtain a baseline for the class as a whole, and we focus primarily on qualitative data from individual interviews to examine students' thinking in greater detail.

Early in the second semester of the course, the students were given a quiz that included two multiple-choice multiple-response questions taken directly from Galley's paper (given originally at the beginning of a physical chemistry course), for comparison. Here, we look at one of those questions:

*An O-P bond in ATP is referred to as a "high-energy phosphate bond" because:*
  *A. The bond is a particularly stable bond.*
  *B. The bond is a relatively weak bond.*
  *C. Breaking the bond releases a significant quantity of energy.*
  *D. A relatively small quantity of energy is required to break the bond.*

Students were instructed to "put the letters corresponding to all the correct answers"; this is apparently different from Galley's students, who were given a limited set of choices ("A and C", "B and C", etc.). In both our class and Galley's class, choices B and D were considered correct. The intent was that students would recognize that energy is released because a relatively weak bond was broken and a relatively strong bond was formed, and that no energy is released by the actual breaking of the bond.[1]

Over the course of the year, the research team conducted 22 semi-structured interviews with 11 students on various topics related to the course. Two of these interviews, with two pre-med students, included explicit discussion of the ATP quiz question. The first interview, with Gregor[2], took place immediately after class on the day that the quizzes were handed back. Gregor brought up the quiz spontaneously in response to a general prompt about the role of biology in the physics course. The second interview, with Wylie, was three weeks later, after the research team had seen the Gregor interview data, so the interviewer prompted Wylie more directly about the ATP quiz question, in order to compare with Gregor.

---

[1] The question, as written, may have been misleading because it asks about the reason for using a term that is itself misleading. Because of this, we believe the question is more valuable as a formative task than as an assessment, and we focus on how the students subsequently thought through the question in interviews. In any case, we have no evidence that this aspect of the question was the source of student confusion; e.g., students had the opportunity to request regrades, and no student requested a regrade on this basis.

[2] All names are pseudonyms.

# RESULTS

On the quiz question, 79% of the class (N=19) selected choice C (breaking the bond releases energy) as a correct answer (whether on its own or along with other answer choices). Our sample size is too small to draw meaningful conclusions from a more detailed breakdown of the quantitative data. We bring this result primarily to show that our class is broadly comparable to Galley's results, in which 87% of students chose C. Galley interprets this as a sign of a "persistent misconception." However, the interview data illustrate that the interpretation must be more complex.

Gregor had selected B, C, and D as correct answers on the quiz, and lost points for choice C. Gregor explains why he chose C (though this retrospective explanation may or may not represent exactly what he was thinking while taking the quiz):

*"I put that when the bond's broken that's energy releasing. Even though I know, if I really think about it, that obviously that's not an energy-releasing mechanism. Because like, you can't break a bond and release energy, like you always need to put energy in, even if it's like a really small amount of energy to break a bond. Yeah, but like. I guess that's the difference between like how a biologist is trained to think, in like a larger context and how physicists just focus on sort of one little thing. Whereas like, so I answered that it releases energy, but it releases energy because when an interaction with other molecules, like water, primarily, and then it creates like an inorganic phosphate molecule that has a lot of resonance. And is much more stable than the original ATP molecule. So like, in the end releases a lot of energy, but it does require like a really small input of energy to break that bond. So I was thinking that larger context of this reaction releases energy. Because I know what the reaction is, ya know? So, um, not, does the bond breaking release energy."*

Gregor demonstrates a sophisticated understanding of the ATP hydrolysis reaction, and makes clear that his justification for choosing C on the quiz (at least in retrospect) does not correspond to the standard "misconception" that bond breaking releases energy. He displays understanding of the intended resolution of the apparent paradox: energy is released not by the breaking of a bond but by the formation of other more stable bonds. In thinking back over the question, Gregor stands by his answer, but also recognizes the correctness of the quiz answer key. He attributes the discrepancy to differing interpretations of what the question is asking (and even assigns this reasoning to

the other students who answered the question the same way):

> "When I was taking the test, I guess I was thinking breaking this bond then leads to these other reactions inevitably. That result in an energy release … I don't [argue] that breaking a bond releases energy, but just like in a larger biological context, that reaction does release energy. So that's what me and apparently like 80% of the class was thinking."

Gregor then, following up on a thread that begins above, ties these differences in perspective to differences between the disciplines:

> "Because I guess like in biology it's not as important to think about like breaking this bond doesn't release energy and then all these other things that happen do release a lot of energy. So, we're, I've just been taught like for a long time that like ATP going to ADP equals like a release of energy. … I guess that's just the difference between physics and chemistry and biology. … It's just your scale. Like, physic[ists] really love to think about things in vacuums, and like without context, in a lot of senses. So, you just think about like whatever small system you're-- isolated system you're looking at, and I guess chemist or biologists thinking about more of like an overall context, that like wherever a reaction or process is happening, like that's important to what's going on."

Gregor and his classmates have biology backgrounds, and their experience talking about ATP and bond breaking is in biology and chemistry courses; now he believes he is seeing a different perspective in a physics course. (Of course, Gregor does not know that this quiz question originated in a chemistry education paper; perhaps his reaction would have been different if he, like Galley's students, had encountered the question in a chemistry course.) He sees the boundary of the phenomenon under consideration as a salient difference between the disciplines. (Gregor's and other students' views on the relationships between the disciplines are explored further in another paper.[8]) In this case, he is not talking about physical scale, but about whether we are looking at the breaking of a bond on its own (which requires an input of energy) or the ATP hydrolysis reaction as a whole (which releases energy).

Wylie also answered B, C, and D on the quiz. Like Gregor, both his multiple-choice responses on the quiz and his responses in the interview were consistent with holding two pictures simultaneously. However, Wylie apparently had not reconciled these two pictures prior to the interview to the same extent that Gregor had. Wylie is aware that he still has reconciliation to do: in thinking back over the question, he says "there's obviously a conflict" between breaking bonds (in ATP) releasing energy and forming bonds (in general) releasing energy. Wylie explains that he answered C because "the result of ATP hydrolysis is ADP, which is much more stable, because I know this from chemistry.… And we have energy released. So obviously you're going from an unstable state to a more stable state." He also justifies his choice of D ("a relatively small quantity of energy is required to break the bond"), "because if something is really unstable, if something is really highly charged, then all it needs is a little push, and that's it, it just goes downhill." Putting it together, Wylie says:

> "If I were to rationalize [the physics professor's] model, then I would have to say ATP breaks down into ADP plus something. There's a bond formed between the phosphate and something that makes it more stable. And this part is the part that releases the energy. … It's not the breaking of the bond that's releasing the energy. Because when, in breaking of the bond, you actually require energy, but the result of the breaking of the bond is that you get energy."

Even though Wylie does this reconciliation to explain why C was deemed incorrect, possibly in real time during the interview, he remains unsatisfied with this conclusion. Like Gregor, Wylie ultimately connects the discrepancy to disciplinary differences:

*Wylie: If … that same question was in a biology course, and I picked C, I would get points.*
*Interviewer: Why do you think that is?*
*Wylie: Because I think in the biology course, the focus of the question would be on the significant quantity of energy, not necessarily breaking the bond. … Breaking the bond in ATP gives you energy. That's what a biologist might think. … But this is more specific. This is going into, you know, exact details.*

Wylie, too, distinguishes between a "biology" approach, in which the focus is on the entire reaction that releases energy, and the "more specific" approach that he associates with the physics class, which focuses on the "exact details."

At the end of the day, Wylie has not gone all the way in building a coherent model and knows that he has further to go:

> "But … I keep thinking that there have to be things that, you know, just like ATP, you know they

*are macromolecules. They're not as stable together, but when they break down they're more stable separately...what do you do with that? How would you release energy in that sense? I don't know. I'm really just kind of unclear on that."*

## DISCUSSION

These case studies illustrate examples of how students' perceptions of the disciplines influence how they reason about cross-disciplinary concepts. Both Wylie and Gregor have come to terms (to varying degrees) with multiple perspectives on the same phenomenon, are able to articulate how their interpretation of a problem depends on the disciplinary context. We do not claim that Wylie and Gregor are typical in this regard. However, these two students provide a framework for thinking about others. To examine the extent to which the rest of the class has reconciled these perspectives, we gave the students a capstone essay question on an exam, which we analyze in a future paper.[9] We found that a substantial portion of the class was able to do this reconciliation given appropriate supports, even if they might not have done so spontaneously.

Novick and Galley both write about the idea of energy released by the breaking of bonds as a "misconception" that needs to be replaced with the correct model. Biology courses are seen as a source of this misconception, which then transfers into students' thinking about chemistry. Teichert & Stacy focus instead on building on students' existing ideas, and using instruction to resolve inconsistencies and integrate new ideas with students' prior intuitive conceptions. Still, the implication is that students will end up with one correct model, and the difference is just in how to get there most effectively. Our data suggest a third approach.

The students' responses to the ATP question combine statements that many instructors would see as mutually inconsistent. This result is not consistent with the view that the students hold unitary "misconceptions." Rather, the data are most consistent with the resources framework [10], in which students possess multiple resources, which can be preferentially activated in a context-dependent way. In particular, our interview data suggest that the disciplinary setting has a key role in determining which resources are activated, and Gregor and Wylie may be unusually self-aware in articulating this.

This is important first of all because disciplinary silos remain present even when interdisciplinary courses do some bridging, and therefore pedagogy and education research around cross-disciplinary topics will need to be sensitive to these issues of disciplinary context. But furthermore, this context-dependent reasoning may even be desirable to some degree. The goals of interdisciplinary education include being able to reason **within** each discipline, using its own native tools, in ways that are informed by and coherent with the other disciplines. We want our students to be able to make choices about how to model a system based on the questions that they are trying to answer. In some circumstances, it is appropriate to consider the individual steps of the ATP hydrolysis reaction mechanism and keep track of which bonds are broken and which bonds are formed, or to track the energy transformations and transfers that take place within this reaction. In other circumstances, the relevant features of the reaction are that ATP is broken into ADP and phosphate and that energy is released, and this relatively black-boxed picture is a useful way to think about the reaction in its larger biological context. Interdisciplinary competency in physics, biology, and chemistry incorporates both of these models.

It is not sufficient to attempt to replace a "misconception" with one domain's scientifically accepted model, but in this vision, it is also not necessarily sufficient to merge the disciplinary perspectives into a single coherent model. Rather, our students are engaged in something productive by seeing that physics may look at a system one way and biology may look at it another way, and by working to reconcile how these perspectives fit together.

## ACKNOWLEDGMENTS

The authors thank the University of Maryland Physics Education Research Group and Biology Education Research Group. This work is supported by the NSF Graduate Research Fellowship (DGE 0750616), NSF-TUES DUE 11-22818, and the HHMI NEXUS grant.